\documentclass[preprint,showpacs,showkeys,preprintnumbers,amsmath,amssymb]{revtex4}
\usepackage {graphicx}

\begin{document}

\title{Analysis and calibration of absorptive images of Bose-Einstein condensate
at non-zero temperatures}

\author{J.~Szczepkowski,$^{1}$
        R.~Gartman,$^{2}$
        M.~Witkowski,$^{3}$
        \L{}.~Tracewski,$^{4}$
        M.~Zawada,$^{2\ast}$
        W.~Gawlik$^{5}$}

\affiliation{
$^{1}$Institute of Physics, Pomeranian Academy, Arciszewskiego 22b, 76-200, S\l{}upsk, Poland,\\
$^{2}$Institute of Physics, Nicolaus Copernicus University, Grudzi\c{a}dzka 5, 87-100 Toru\'n, Poland,\\
$^{3}$Institute of Physics, University of Opole, Oleska 48, 45-052 Opole, Poland,\\
$^{4}$Institute of Experimental Physics, University of Wroclaw, Plac Maksa Borna 9, 50-204 Wroc\l{}aw, Poland,\\
$^{5}$Institute of Physics, Jagiellonian University, Reymonta 4, 30-059 Krak\'ow, Poland.\\}

\email{zawada@fizyka.umk.pl} 


\begin{abstract}
We describe the method allowing quantitative
interpretation of absorptive images of mixtures of BEC and thermal
atoms which reduces possible systematic errors associated with
evaluation of the contribution of each fraction. By using known
temperature dependence of the BEC fraction, the analysis allows
precise calibration of the fitting results. The developed method
is verified in two different measurements and
compares well with theoretical calculations and with
measurements performed by another group.
\end{abstract}

{\it Submitted to: Rev. Sci. Inst.} 


\maketitle

\section{Introduction}

Analysis of phase transitions offers valuable data of many physical systems. This is particularly true
for studies of Bose-Einstein (BE) condensation in diluted gases
\cite{Cor95,Ket95}. While such studies are very important
scientifically, they pose many experimental challenges. The main
difficulty is caused by extremely low temperatures in which
Bose-Einstein condensates (BECs) are created and investigated, on
the order of 100 nK. It is therefore essential to develop reliable
detection/imaging methods for ultracold atoms close to the
phase-transition point (density and temperature). Several such
methods have been developed by many groups both destructive, like
absorptive imaging, and nondestructive, like phase
contrast \cite{Ket96} and polarization imaging \cite{Hul97}. The
simplest and most often applied detection technique is the
absorptive imaging and this is the method on which we concentrate
in this paper.

In the absorptive imaging one records images of the shadow cast
onto a camera sensor by atoms usually released from a trap during their
free gravitational fall.
This yields 2D distribution of optical density which reflects
spatial density profile of the atomic cloud. Analysis of such
profiles allows derivation of all relevant physical parameters of
the investigated sample. The main difficulty in such analysis is
associated with the fact that at finite temperatures the BEC
fraction is always associated with some fraction of thermal
(non-condensed) atoms. The thermal fraction plays a very important
role in the data analysis as it allows determination of the cloud
temperature. This fraction decreases with the falling temperature
of the cloud. Each of the fractions has different density
distribution and contributes differently to the recorded image.
The coexistence and overlap of the two fractions results in a
bimodal distribution of the optical density which raises
interpretational problems. Below, we show that a
simplistic analysis of such bimodal distributions by fitting them
to a sum of the Gaussian and Thomas-Fermi
functions corresponding to the thermal and condensate
fractions, respectively, is not satisfactory and leads to
systematic errors.

The problems associated with the analysis of absorption profiles
are not new and were already noticed by several other groups
\cite{Ket99,Cor96,KurnPhD}. One attempt which partly avoids the
problem of the bimodal distribution is the spatial separation of
thermal and BEC fractions by the Bragg diffraction
\cite{Kurn99,Ger04} or by application of an optical lattice
\cite{Fer02}. These methods, however, are neither easy nor ideal
as they also can introduce inter-fraction interaction. So far, the
best known method is to analyze the thermal fraction only in its
outer regions, outside the degenerate regime, where it is possible
to use a simple classical description \cite{Ket99}. The
main disadvantage of this method is that it is not easy and rather
arbitrary to find the correct size of the excluded central region.
As this region is enlarged, the systematic error of the
fitted parameters is reduced, but the S/N ratio
decreases. On the other hand, if the excluded central
region is too small, the fit is performed to the data which sample
also the edges of the degenerate region. The
resulting systematic errors can be minimized, for
example by applying corrections based on a numerical solution of
the ideal Bose-Einstein distribution \cite{Cor96}, which is not a
trivial task.

The present paper introduces the method allowing
quantitative analysis of the absorptive pictures
which
ensures correctness of the size of the excluded
degenerated region. We present the algorithm for analysis of the
bimodal distributions which yields accurate ratio of the BEC and
thermal fractions at finite temperatures. The analysis allows
calibration of the thermal fraction fits and minimizes
number of measurements necessary to obtain statistically
meaningful averages.

Section~\ref{sec:method} presents the procedure of fitting the
bimodal distribution of optical density and the method for the fit
optimization. In Section~\ref{sec:comp} we compare our method
with other commonly used approaches. In
Section~\ref{sec:results} we present examples of analysis of the
experimental data with the two methods, the simplest one
which uses a sum of a Gaussian and Thomas-Fermi
distributions and the one we have developed. The paper is
concluded in Section~\ref{sec:conc}.

\section{Method for analyzing the images of a condensate in non-zero temperature \label{sec:method}}

This Section describes the main points of our method for
analyzing the BEC pictures and its calibration.

\subsection{Fitting to the bimodal distribution \label{sec:fit}}

\begin{figure}
\begin{center}
      \resizebox{0.3\columnwidth}{!}{\includegraphics{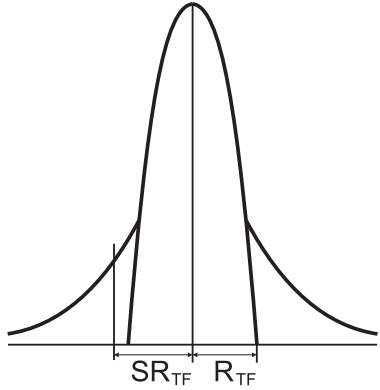}}
      \caption{Illustration of the spatial extension of the
      thermal (outer) and BEC (central) fractions fitted by different distributions. The
      scaling factor $S$ is chosen for each picture individually to
      account for the distortion of the density distribution in the
      intermediate region, as described in Subsec. \ref{sec:cal}}.
      \label{fig:gft_mod}
\end{center}
\end{figure}

Two-dimensional picture of a column optical density (OD) contains
information on the spatial distribution of the column atomic
density in a cloud
$\tilde{n}\left(r,z\right)=OD\left(r,z\right)/\sigma_{0}$, where
$r, z$ are the radial and axial coordinates,
respectively, (Fig.\ \ref{fig:gft_mod}) and  $\sigma_{0}=3\lambda^2/2\pi$
is the normalized cross-section for atomic absorption at
wavelength $\lambda$. By column densities we understand regular
densities integrated over the local sample thickness, i.e. the
column length. From the OD distribution recorded with light
intensity $I$, the non-saturated distribution $OD_{n}$
can be calculated as $OD_{n}=OD+\left( 1-\exp(-OD) \right)
I/I_{sat}$, where $I_{sat}$~is the~saturation intensity for the
imaging transition. If the expansion of the cloud is
small, we have to take into account
that the cloud can be completely dark for the absorption
probe beam (see, e.g. \cite{Lew03}).

Well above the critical temperature, the density
distribution in the thermal cloud can be described with the
classical Boltzmann distribution. The column density is described
then by the Gaussian function:

\begin{equation}
OD_{Gauss}(r,z)=OD_{Gpeak}\ \mathrm{exp}\left[ -\frac{1}{2}\left( \frac{r-r_c}{%
\sigma _{r}}\right) ^{2}-\frac{1}{2}\left( \frac{z-z_c}{\sigma _{z}}\right)
^{2}\right] ,  \label{eq:fGauss}
\end{equation}

\noindent with $\sigma _{r},\sigma _{z}$ being the half-width of
the atomic density distribution in the radial and axial
directions, respectively, $OD_{Gpeak}$ denotes the maximum value
of the thermal fraction density, and $(r_c,z_c)$ are spatial
coordinates of the maximum. For temperatures close to and lower
than the critical value, the density distribution becomes
predominantly the Bose distribution. Then, if the
chemical potential is set to zero, the column optical density can
be described by the, so called, Bose-enhanced Gaussian function
\cite{Ket99, KurnPhD}:

 \begin{equation}
         OD_{EnhGauss}(r,z)= OD_{Gpeak}\frac {g_{2}\left[\mathrm{exp}\left[ -\frac{1}{2}\left( \frac{r-r_c}{\sigma _{r}}\right) ^{2}-\frac{1}{2}\left( \frac{z-z_c}{\sigma _{z}}\right)
^{2}\right]\right]}{g_{2}(1)},
        \label{eq:fEGauss}
 \end{equation}

\noindent where $g_{2}\left(x\right)=\sum^{\infty}_{n=1} \left(x^n\right)/ \left(n^{2}\right)$ (see, e.g. \cite{Hua87}).

With increasing distance from the position of
the maximum density, the series terms in numerator of
(\ref{eq:fEGauss}) decrease to zero. At appropriate distance,
function (\ref{eq:fEGauss}) becomes the Gauss function
(\ref{eq:fGauss}) which justifies description of the density
distribution at the edges of the thermal fraction by
function (\ref{eq:fGauss}). Nevertheless, more accurate results
are obtained if the first three terms of series
(\ref{eq:fEGauss}) are used instead. Including of yet more
terms only increases computation time without noticeably improving the
accuracy.

In the BEC fraction, on the other hand, the distribution
of the column optical density of atoms in the Thomas-Fermi
regime can be described by the TF profile, a clipped parabola, 

\begin{equation}
OD_{TF}(r,z)=OD_{TFpeak}\ \mathrm{max}\left[0, \left( 1-\left( \frac{r-r_c}{R_{r}}%
\right) ^{2}-\left( \frac{z-z_c}{R_{z}}\right) ^{2}\right) ^{3/2}\right] ,
\label{fTF}
\end{equation}

\noindent where $R_{z},R_{r}$ are the TF radii in the radial and
axial directions, respectively, and $OD_{TFpeak}$ denotes
maximum of the condensate optical density. When a condensate is
not in the TF regime, the density distribution is well
approximated by a Gauss function \cite{Ket99}.

All pictures in our procedure are taken with the
condensates that were expanding for time $t$ after their releasing
from MT. Knowing spatial distribution of the atomic
density in the falling cloud, the initial atomic temperature can
be determined \cite{Ger04_t}:

\begin{equation}
T=\frac{2\tau_{r}^2}{1+3\tau_{r}^2}T_{r}+\frac{1+\tau_{z}^2}{1+3\tau_{z}^2}T_{z},
\end{equation}

\noindent where $\tau_{i}=\omega_{i}t$   for $i=z,r$ and
$T_i=(m/2 k_B)\left[\omega_i^2 \sigma_i^2(t) /(1+\omega_i^2 t^2)\right]$  are the effective
temperatures measured after expansion. Moreover, the
distribution allows determination of the numbers of atoms in each
fraction, the thermal fraction $N_{th}$ and the condensate
fraction $N_{BEC}$:
\begin{equation}
N_{th}=\left(2\pi\right)^{3/2} \frac{OD_{Gpeak}}{\sigma_0}
\sigma_{r}^2(t)\sigma_{z}(t),
\end{equation}

\begin{equation}
N_{BEC}=\frac{8}{15}\pi \frac{OD_{TFpeak}}{\sigma_0} R_{r}^{2}(t) R_{z}(t).
\end{equation}

The values of the distribution parameters are to be derived by
fitting the $OD$ distribution functions to the experimentally
recorded profiles. For pictures corresponding to homogenous
samples consisting exclusively of either thermal atoms or pure
BEC, the functions (\ref{eq:fGauss}), (\ref{eq:fEGauss}) or
(\ref{fTF}) can be fitted, as appropriate. Such fits are performed
for the radial and axial sections independently. The sections
cross the center of mass which coincides with the maximum of
optical density. In the least-squares fitting procedure the {\tt
MINUIT} library \cite{minuit} was used.

In a bimodal atom cloud, containing both the BEC and thermal
fractions, the recorded pictures consist of two regions, the
external region occupied by the thermal cloud only and the
internal one where the two fractions coexist. In the
later region and close to the border between the two regions the
density distribution is distorted by the interaction between the
fractions and by the Bose enhancement of the thermal fraction
\cite{Ket99, KurnPhD}. To reduce the effect of this distortion,
the fitting has to be performed in several steps.

The first step is to determine the region occupied by the
condensate and its direct neighborhood. For this sake, we
approximate the bimodal distribution by the sum of functions
(\ref{eq:fGauss}) and (\ref{fTF}) with some offset and fit it to
the column density picture of the atomic cloud. The fit is
performed for each direction independently with the least-squares
fitting procedure based on the {\tt MINUIT} library.

Next, using the parameters derived from the fitting curve
(\ref{fTF}), the initial BEC extension, i.e. the TF radii, $R_z,
R_r$, are determined. This allows to subtract the BEC contribution
from the analyzed picture. To account for distortions in
the intermediate region, the size of the subtracted area is taken
with some margin such that the area dimensions are bigger than
$R_z$ and $R_r$ by scaling factor $S$ (see
Fig.~\ref{fig:gft_mod}). The procedure of exact determination of
the $S$ value is described in the following subsection.

After removing the BEC contribution with appropriate
safety margins determined by the scaling factor, the remaining
image consists already of a pure thermal fraction and can be
fitted by first three terms of series (\ref{eq:fEGauss})
with some
background. We do it with the least-squares method by the
2D NonLinearFit function in Mathematica 5.1. The fit
parameters allow calculation of the atom number in a thermal
cloud and its temperature.

Having determined the optical density distribution in the thermal
fraction and the background level, they can be subtracted from the
initial picture with full bimodal distribution.
Additionally, at this stage data points which are below
5\% threshold are rejected to eliminate the contribution of the
intermediate, distorted region of the cloud picture.

To such evaluated data the TF profile (\ref{fTF}) is fit by the 2D
least-squares method, which yields atom number in the BEC fraction
and the appropriate TF radii.  For  minimization of
$\chi^2$, the NMinimize function in Mathematica is used.

\subsection{Calibration of the thermal fraction region \label{sec:cal}}

The value of coefficient $S$ affects the calculated temperature
and atom number of the BEC fraction in a bimodal distribution.
Taking too big margin, i.e. too big $S$, eliminates too large
region which contains information on the density distribution of
the thermal fraction, thereby lowering the signal/noise ratio and
reducing the fit accuracy. Too small value of $S$, on the other
hand, introduces systematic errors by including the distorted
regions.

\begin{figure} \begin{center}
      \resizebox{0.4\columnwidth}{!}{\includegraphics{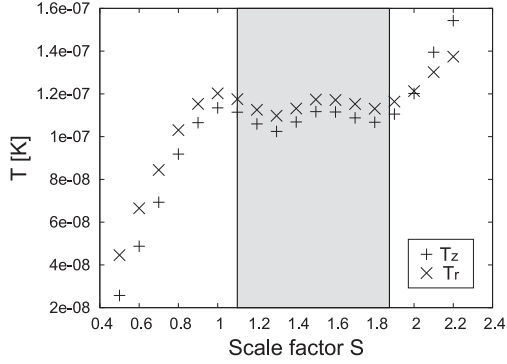}}
      \caption{Temperature values in the radial and axial directions determined for experimental recording analyzed with different values of $S$. The marked region indicates the range of $S$ values for which the determined temperature does not change by more than one standard deviation.}       \label{fig:TS} \end{center} \end{figure}

Fig.\ \ref{fig:TS} illustrates the effect of the value of scaling
factor $S$ on the calculated temperature of the bimodal cloud in
the radial and axial directions. All fits used for this figure
were performed for the same picture of a thermal condensate.
Fig.\ \ref{fig:TS} illustrates that there exists a fairly wide range
of the $S$ values where the determined temperature does not change
by more than one standard deviation (shaded range in Fig.\ \ref{fig:TS}). For $S<1$, the determined temperature is
underestimated by including the intermediate border region
affected by the BEC fraction, while for $S>1.8$, the temperature
is not correctly estimated because of low S/N
ratio of a too much reduced picture. The choice of appropriate
value of $S$ requires taking into account also the effect of $S$
on the number of atoms in the BEC fraction derived from the fit.
This effect can be seen when studying the phase-transition plot,
i.e. the dependence of the BEC size normalized to all atoms in a
bimodal cloud, $N_{0}/N=N_{BEC}/(N_{BEC}+N_{th})$, on the reduced
temperature $T/T_C(N)$ with
$T_C(N)=\hbar/k_B\left[N\cdot(\omega_r^2
\omega_z/1.202)\right]^{1/3}$ being the critical temperature.

\begin{figure}
\begin{center}
      \resizebox{0.4\columnwidth}{!}{\includegraphics{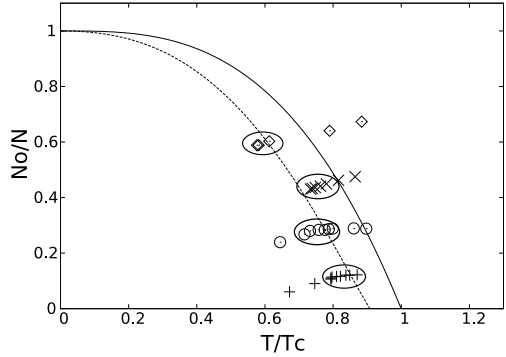}}
      \caption{
Condensate fraction versus normalized temperature. The points
represent results of fitting the bimodal distributions obtained
with different values of scaling parameter S (between 0.8
and 1.8) to four different images (marked by different symbols:
$\diamondsuit, \bigcirc, +$, and $\times$), as described in
text. The elliptical contours depict the regions of
concentration points, i.e. those where the derived values of atom
number and reduced temperature weakly depend on $S$. The
solid line represents function $N_{0}/N=1-(T/T_C(N))^3$
\cite{Gro50} and the broken line represents behavior of a
trapped, semi-ideal Bose gas \cite{Nar98}. }
      \label{fig:nt}
\end{center}
\end{figure}

Nonetheless, the check that the measured temperature of
the cloud is independent of the size of the exclusion region is
not the sufficient criterion of the fit quality.
Particularly, if the measured data is heavily affected by
noise or if the image sizes of either fraction are comparable with
the image resolution, the temperature stability region
(shaded area in Fig.\ \ref{fig:TS}) can be very narrow or
even vanish completely. We have, therefore,
studied further consequences of various choices of the $S$
values.

In Fig.\ \ref{fig:nt} we depict number of atoms in the BEC
fraction versus the reduced temperature for four typical images of
the bimodal cloud. Bimodal distributions corresponding to
different values of $S$ were fitted to each of the pictures as
described above (Subsec. \ref{sec:fit}).  For each recorded image
the fitting procedure was performed for different values of $S$
which  were increasing by constant increments from 0.8 to 1.8. For
a specific range of $S$ the derived atom numbers, as well as the
temperatures, concentrate around specific values, the
"concentration points". They change very little with $S$ by no
more than $\pm5$\%. This fact indicates that $S$ values within the
given range provide optimal separation of the perturbed thermal
fraction form the non-perturbed one which allows proper
description of the thermal cloud by function (\ref{eq:fGauss})
without sacrificing the S/N ratio too much.

A very convincing verification of the fit quality is the
position of the "concentration points" on the phase transition
plots relative theoretical curves, like in Fig. \ref{fig:nt}
\cite{Gro50,Nar98}. If a given "concentration point"
appears far from the theoretical curve, it indicates that the
corresponding image was too much affected by some
nonstatistical noise, e.g. caused by interference fringes
or systematics. In such case, another fitting should be
tried with another (bigger or smaller) background around
the atomic cloud. Our experience shows that in
about 90\% of all cases a single repetition of the
fitting procedure yields a good result. The remaining 10\% is most
often associated with systematic errors.

\begin{figure}
\begin{center}
      \resizebox{0.4\columnwidth}{!}{\includegraphics{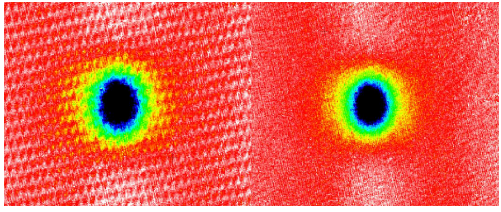}}
      \caption{Examples of the heavily fringed and defringed images.
      The interference fringes have been removed from the original
      image as described in text.
}
      \label{fig:fr}
\end{center}
\end{figure}

Interference fringes can be removed to a large
extent from the image by a sequential subjecting the data to FFT,
the mask corresponding to the fringe frequencies and to
reverse-FFT. This can be done, e.g. with the ImageJ software
\cite{fringes}. Examples of the heavy fringed and
defringed images are presented in Fig. \ref{fig:fr}.

Under conditions of our experiment, the optimum $S$
values are between $1.1$ and
 $1.4$ and depend on the size of the BEC fraction.
The center of the "concentration point" on the phase transition
plots gives the correct value of S, i.e. the correct size of the
excluded degenerated region, while the size of the "concentration
point" allows to estimate its statistical uncertainty.

\section{Comparison to other methods \label{sec:comp}}

In order to compare our method with the other
approaches we have analyzed a condensate image
created by a computer simulation. The simulation created a
2D atomic density distribution of a bimodal cloud with 0.4 BEC
fraction in a given trapping potential. It was based on modelling
of the BEC part by the Castin and Dum theory \cite{Cas96}, the
thermal fraction by the Bose-enhanced distribution (2), and by
taking into account their ballistic expansion within 22 ms.
Finally, a noise typical for the recorded absorption
pictures was added to such constructed simulation.
In our simulation the thermal component is treated as an ideal Bose gas, while the condensate part is assumed to be in the TF regime.

\begin{figure}
\begin{center}
      \resizebox{0.4\columnwidth}{!}{\includegraphics{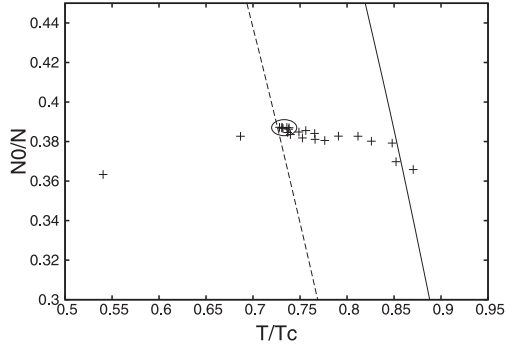}}
      \caption{
Expanded dependence of the condensate fraction versus
normalized temperature. The points represent results of fitting
the bimodal distributions to the image generated by simulation, as
described in text, with the $S$ values in the range
between 0.8 and 3.0. The elliptical contour marks the regions of
concentration of nine points, where the derived values of
the atom number and reduced temperature weakly depend on
$S$. The solid line represents function
$N_{0}/N=1-(T/T_C(N))^3$ \cite{Gro50}, the broken line
represents behavior of a trapped, semi-ideal Bose gas
\cite{Nar98}. }
      \label{fig:simN}
\end{center}
\end{figure}

In Fig.\ \ref{fig:simN} we depict the number of atoms in
the BEC fraction versus the reduced temperature obtained by
fitting the bimodal distributions to the image generated by the
simulation. The $S$ values used in the fits are in the range
between 0.8 and 3.0 with a step of 0.1. Similarly as in
Fig.\ \ref{fig:nt}, the elliptical contour, indicates the regions
of concentration of nine points, where the derived values of atom
number and reduced temperature weakly depend on $S$.

The basic results of our analysis, i.e. the number of atoms in
both the condensate $N_0$, and thermal
$N_{th}$, components, temperature $T$ and the
Thomas-Fermi radii $R_z, R_r,$ are compared with the
values preset in the simulation and generated by other
methods. The first method, due to its simplicity being probably
the most common, fits a sum of the 2D Gaussian
(\ref{eq:fGauss}) and TF (\ref{fTF}) distributions to the
experimental data. In the second method the Gaussian
function is fitted to the wings of the spatial
distribution, then subtracted from the whole distribution and
the remaining data is eventually
fitted by the TF function. Fitting a 2D Gaussian to the wings of
the distribution was widely used in the early
experiments on BEC (e.g. \cite{Cor96,Kas99,Ing01,Gri02}).
The noise added to the simulated picture
reproduces real experimental conditions and causes that
neither of the methods gives the perfect fit.
Still, the described method provides the the closest agreement
with the simulation parameters.

\begin{figure}
\begin{center}
      \resizebox{0.4\columnwidth}{!}{\includegraphics{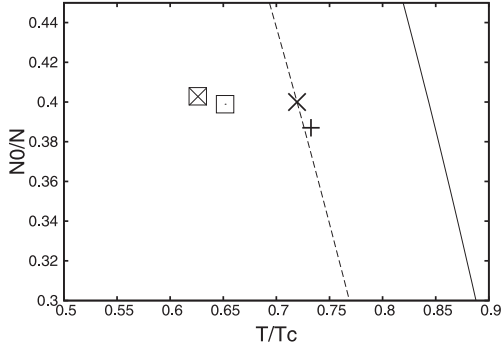}}
      \caption{
Condensate fraction versus normalized temperature -
the results generated by different methods of image analysis
applied to the simulated picture of a bimodal distribution. The
value used in the simulation is depicted as $\times$. Our method
gives the value represented by $+$, a simple 2D sum of
the Gaussian and TF distributions gives $\boxtimes$, while a Gauss
function used instead of the Bose-enhanced Gaussian gives
$\boxdot$.
The solid and broken lines have the same meaning as in
Fig.\ \ref{fig:simN}.}
      \label{fig:comp}
\end{center}
\end{figure}

\begin{table}
    \centering
        \begin{tabular}{||c||c|c|c|c|c||}
        \hline
~   & $N_0$ & $N_{th}$ & $T$ & $R_z$ & $R_r$  \\
\hline
Simulation & 1 & 1 & 1 & 1 & 1\\
Our method & 1.0104 & 1.0660 & 1.0304 & 1.0045 & 0.995 \\
2D SUM & 1.0199 & 1.0082 & 0.8737 & 1.0081 & 0.995 \\
GTF & 1.0264 & 1.0310 & 0.9143 & 1.0084 & 0.996 \\
\hline
        \end{tabular}
    \caption{Results of the analysis of the simulated image
by three different methods: our method, "2D SUM" - the method with
fitting a 2D sum of the Gaussian (\ref{eq:fGauss}) and TF
(\ref{fTF}) distributions, "GTF" - the fitting of the Gaussian
function to the wings of the distribution and the TF function to
the condensate part.
}
    \label{tab:Comparison}
\end{table}

\section{Typical examples\label{sec:results}}

In this Section we discuss results of the analysis of the
typical experimental images obtained with our
setup \cite{Byl08}. The experiment was devoted to studies
of the free-fall dynamics of a finite-temperature condensate of
$^{87}$Rb in the F=2 state and is described in more detail
elsewhere \cite{Zaw08}. Here, we present two examples showing how the
results derived with the method described in Sec.\ \ref{sec:method} 
compare with those obtained with the simple "2D
SUM" method based on summation of the Gauss and TF distributions.
This comparison well illustrates the potential of our method.

\subsection{Dependence of $N_{0}/N$ on $T/T_c$}

\begin{figure}
\begin{center}
\resizebox{0.8\columnwidth}{!}{\includegraphics{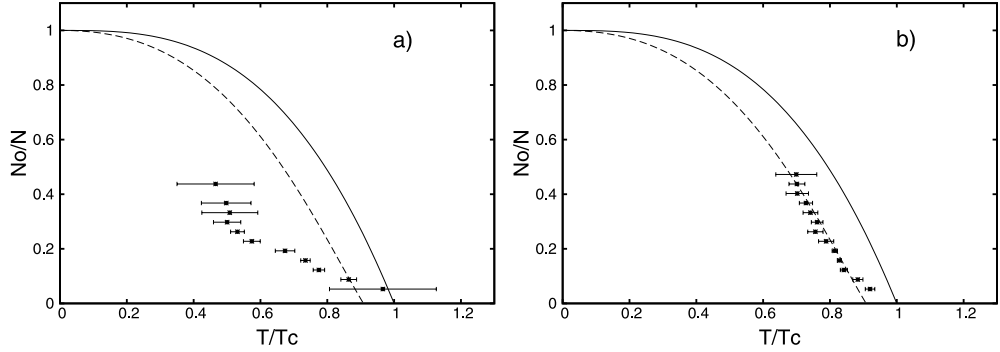}}
\caption{Comparison of the typical experimental dependences of
      the BEC fraction, $N_0/N,$ on the reduced temperature, $T/T_C(N),$
      yielded by 210 images. The points in (a) result from
      simple fitting the 2D sum of functions (\ref{eq:fGauss}) and (\ref{fTF})
      to the sections of absorption images, whereas
      the points in (b) correspond to the fitting procedure
      described in this paper. The solid and broken lines are the
      same as in Figs.\ \ref{fig:nt}, \ref{fig:simN}, and \ref{fig:comp}.}
      \label{fig:gtf}
\end{center}
\end{figure}

Fig.\ \ref{fig:gtf} represents a typical experimental dependence of
the BEC fraction $N_0/N$ on the reduced temperature, $T/T_C$,
analyzed with two approaches. The points represented in
Fig.\ \ref{fig:gtf}(a) are obtained by using the 2D fitting of the
sum of functions (\ref{eq:fGauss}) and (\ref{fTF}) to the
sections of absorptive images. The points in Fig.\ \ref{fig:gtf}(b)
are obtained using our fitting procedure (Sec. \ref{sec:method}).
As before, the solid lines are the $N_0/N=1-(T/Tc(N))^3$
functions, while the broken ones represent results of the
calculations along the lines of Ref. \cite{Nar98}. According to
Ref.\ \cite{Ket99}, the bigger is the BEC fraction in the
sample, the more the results of a simplistic fit
with a sum of functions (\ref{eq:fGauss}) and (\ref{fTF}) deviate
from the real temperature. The experimental data points are
obtained from relatively small number of 210 BEC images taken at
different temperatures.
As can be seen in Fig.\ \ref{fig:gtf}(b), experimental
points evaluated with our method are in excellent agreement with
the model of Ref.\ \cite{Nar98}, while those shown in Fig.\ \ref{fig:gtf}
(a) deviate dramatically from theoretical
predictions.

\subsection{Temperature dependence of the aspect ratio $R_r/R_z$ of a free falling BEC}

\begin{figure}
\begin{center}
\resizebox{0.8\columnwidth}{!}{\includegraphics{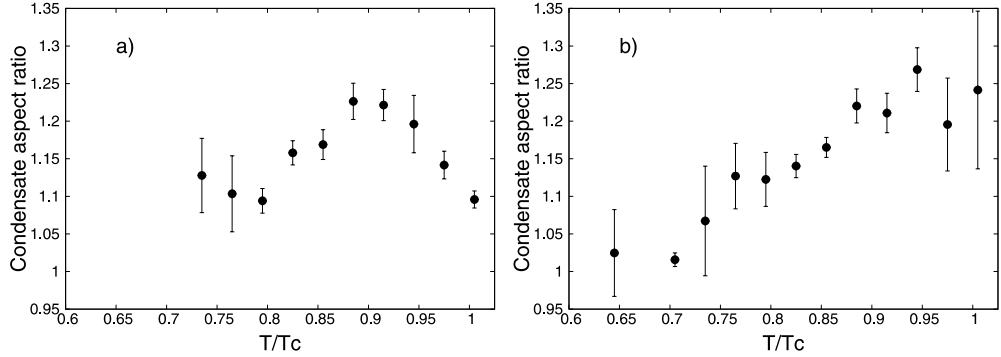}} 
      \caption{Comparison of the typical dependences of the BEC aspect
ratio on the normalized temperature, $T/T_C$, obtained from 150
images taken after $t=15$~ms free fall. (a) depicts results of 2D
fitting of a sum of functions (\ref{eq:fGauss}) and (\ref{fTF})
to the sections of absorption images, points in (b) are
obtained by application of the described procedure.}
      \label{fig:AR}
\end{center}
\end{figure}

Fig.\ \ref{fig:AR} presents results of our measurements of the BEC
aspect ratio as a function of the reduced temperature,
$T/T_C$, evaluated from 150 images of BEC taken at
different temperatures after $t=15$~ms free fall.
As described previously, the data points in Fig.\ \ref{fig:AR}(a)
were obtained by 2D fitting of a sum of functions
(\ref{eq:fGauss}) and (\ref{fTF}) to the sections of
absorptive images whereas those in Fig.\ \ref{fig:AR}(b) by using
our new fitting method. The points evaluated with the new method
behave qualitatively in the same way as in a similar
experiment of Gerbier  $et$ $al.$ \cite{Ger04}. However, in Ref.
\cite{Ger04} the thermal and BEC fractions were completely
separated spatially by Bragg diffraction which eliminated problems
of their proper identification in the absorptive images. Despite
different methods of the fraction separation, our method
yields qualitatively similar results \cite{comparison}. On the
other hand, the points those obtained with the
simple "2D-SUM" method exhibit distinctly
different, nonphysical behavior .

\section{Conclusions \label{sec:conc}}

We have developed the method allowing proper interpretation of
absorptive images of mixtures of BEC and thermal atoms which
reduces possible systematic errors arising from  non-Gaussian distribution of ultra-cold thermal atoms.

The developed algorithm is based on the fitting procedure of 2D
density distributions to the absorption profiles describing the
thermal fraction. By using the well known temperature dependence
of the BEC fraction, the analysis allows precise calibration of
the fitting results and, consequently, reduces number of
measurements necessary to obtain statistically meaningful average
values. We compare our method with the others commonly used. We
have performed experiments verifying the developed method in two
different measurements. Comparison of the results analyzed with
our method and with the simplest fitting procedure demonstrates
that the described method yields far better accuracy and is less
prone to systematic errors. The results interpreted with our
approach are also consistent with theoretical calculations and 
with the results of measurements performed by another group.

\section*{Acknowledgments}

This work has been performed in KL~FAMO, the National
Laboratory of AMO Physics in Toru\'n and supported by the Polish
Ministry of Science. The authors are grateful to J.
Zachorowski for numerous discussions. J.S. acknowledges also
partial support of~the~Pomeranian University (projects numbers
BW/8/1230/08 and BW/8/1295/08).



\begin{thebibliography}{99}


\bibitem{Cor95} M.H.~Anderson, J.R.~Ensher, M.R.~Matthews, C.E.~Wieman, E.A.~Cornell, "Observation of Bose-Einstein Condensation in a Dilute Atomic Vapor," Science {\bf 269}, 198, (1995).
\bibitem{Ket95} K.B.~Davis, M.O.~Mewes, M.R.~Andrews, N.J.~van Druten, D.S.~Durfee, D.M.~Kurn and W.~Ketterle, "Bose-Einstein Condensation in a Gas of Sodium Atoms," Phys. Rev. Lett. {\bf 75}, 3969, (1995).
\bibitem{Ket96} M.R.~Andrews, M.O.~Mewes, N.J.~van Druten, D.S.~Durfee, D.M.~Kurn and W.~Ketterle, "Direct, Nondestructive Observation of a Bose Condensate," Science {\bf 273}, 84, (1996).
\bibitem {Hul97} C.C.~Bradley, C.A.~Sackett and R.G.~Hulet, "Bose-Einstein Condensation of Lithium: Observation of Limited Condensate Number," Phys. Rev. Lett. {\bf 78}, 985, (1997).
\bibitem{Ket99} W.~Ketterle, D.~S.~Durfee, and D.~M.~Stamper-Kurn, {\it Proceedings of the International School of Physics "Enrico Fermi", Course CXL}, edited by M. Inguscio, S. Stringari and C.E. Wieman (IOS Press, Amsterdam, 1999) pp. 67-176.
\bibitem{Cor96} J.R.~Ensher, D.S.~Jin, M.R.~Matthews, C.E.~Wieman and E.A.~Cornell, "Bose-Einstein Condensation in a Dilute Gas: Measurement of Energy and Ground-State Occupation," Phys. Rev. Lett. {\bf 77}, 4984 (1996).
\bibitem{KurnPhD} D.M.~Stamper-Kurn, {\it Peeking and poking at a new quantum fluid: Studies of gaseous Bose-Einstein condensates in magnetic and optical traps }, PhD Thesis, Massachusetts Institute of Technology, (2000).
\bibitem{Kurn99} J.~Stenger, S.~Inouye, A.P.~Chikkatur, D.M.~Stamper-Kurn, D.E.~Pritchard, W.~Ketterle, "Bragg Spectroscopy of a Bose-Einstein Condensate," Phys. Rev. Lett.  {\bf 82}, 4569 (1999).
\bibitem{Ger04} F.~Gerbier, J.~H.~Thywissen, S.~Richard, M.~Hugbart, P.~Bouyer, and A.~Aspect, "Experimental study of the thermodynamics of an interacting trapped Bose-Einstein condensed gas," Phys. Rev. A {\bf 70}, 013607 (2004).
\bibitem{Fer02} F.~Ferlaino, P.~Maddaloni, S.~Burger, F.~S.~Cataliotti, C.~Fort, M.~Modugno, and M.~Inguscio, "Dynamics of a Bose-Einstein condensate at finite temperature in an atom-optical coherence filter," Phys Rev. A {\bf 66}, 011604(R) (2002).
\bibitem {Lew03} H.~J.~Lewandowski, D.~M.~Harber, D.~L.~Whitaker, and E.~A.~Cornell, "Simplified System for Creating a Bose-Einstein Condensate", J. Low Temp. Phys. {\bf 132}, 309, (2003)
\bibitem{Hua87} K.~Huang, {\it Statistical Mechanics, Second Edition}, John Wiley and Sons, New York (1987).
\bibitem{Ger04_t} F.~Gerbier, J.~H.~Thywissen, S.~Richard, M.~Hugbart, P.~Bouyer, and A.~Aspect, "Critical Temperature of a Trapped, Weakly Interacting Bose Gas," Phys. Rev. Lett. {\bf 92}, 030405 (2004).
\bibitem{minuit} F.~James, {\it {\tt MINUIT}, Function Minimization and Error Analysis, Reference Manual Version 94.1}, CERN Program Library Long Writeup {\bf D506}, CERN, (1994)
\bibitem{Gro50} S.~R.~de~Groot, G.~J.~Hooyman, and C.~A.~ten~Seldam, "On the Bose-Einstein condensation," Proc. R. Soc. London A {\bf 203}, 266 (1950).
\bibitem{Nar98} M.~Naraschewski, D.~M.~Stamper-Kurn, "Analytical description of a trapped semi-ideal Bose gas at finite temperature," Phys. Rev. A {\bf 58}, 2423 (1998).
\bibitem{fringes} M.~D.~Abramoff, P.~J.~Magelhaes, S.~J.~Ram, "Image Processing with ImageJ,"
Biophotonics International {\bf 11}, 36, (2004)
\bibitem{Cas96} Y.~Castin and R.~Dum, "Bose-Einstein Condensates in Time Dependent Traps", Phys. Rev. Lett. {\bf 77}, 5315, (1996)
\bibitem{Kas99} B.~P.~Anderson and M.~A.~Kasevich, "Spatial observation of Bose-Einstein condensation of ${}^{87}$Rb in a confining potential", Phys. Rev. A, {\bf 59}, R938, (1999)
\bibitem{Ing01} G.~Modugno, G.~Ferrari, G.~Roati, R.~J.~Brecha, A.~Simoni, M.~Inguscio, "Bose-Einstein Condensation of Potassium Atoms by Sympathetic Cooling", Science, {\bf 294}, 1320 (2001)
\bibitem{Gri02} T~Weber, J~Herbig, M~Mark, H~-C~N\"agerl, and R~Grimm, "Bose-Einstein Condensation of Cesium", Science Express, 10.1126/science.1079699, (2002).
\bibitem{Byl08} F.~Bylicki, W.~Gawlik, W.~Jastrz\c{e}bski, A.~Noga, J.~Szczepkowski, M.~Witkowski. J.~Zachorowski, M.~Zawada, "Studies of the Hydrodynamic Properties of Bose-Einstein Condensate of  ${}^{87}$Rb Atoms in a Magnetic Trap," Acta Phys. Pol. A {\bf 113}, 691 (2008).
\bibitem{Zaw08} M.~Zawada, R.~Abdoul, J.~Chwede\'nczuk, R.~Gartman, J.~Szczepkowski, \L{}.~Tracewski, M.~Witkowski, W.~Gawlik, "Free-fall expansion of finite-temperature Bose-Einstein condensed gas in the non Thomas-Fermi regime", accepted in J. Phys. B.; arXiv:0811.1672 (2008)
\bibitem{comparison} No quantitative comparison of the results of our work and that of ref. \cite{Ger04} is possible because of different conditions of the two experiments. In particular, in our case the free-fall time is $t=15$~ms, while in \cite{Ger04} it was $22.3$~ms. Also, our temperature $T$ is normalized to $T_C(N)$ which depends on the total number of atoms $N$, while in \cite{Ger04} the normalization was to the critical temperature of an ideal gas in a thermodynamic limit.


\end{thebibliography}
\end{document}